\documentclass[a4paper,twocolumn,superscriptaddress,prl,floatfix]{revtex4}
\usepackage{epsfig}
\usepackage{epstopdf}
\usepackage{graphicx}
\usepackage{amsfonts,amsmath,latexsym,amssymb}

\begin{document}

\def\Dy227{Dy$_2$Ti$_2$O$_7$}

\title{Unconventional magnetization processes and thermal runaway in spin-ice Dy$_2$Ti$_2$O$_7$}

\author{D. Slobinsky}

\affiliation{SUPA, School of Physics and Astronomy, University of St Andrews, St Andrews KY16\ 9SS, United Kingdom}

\author{C. Castelnovo}

\affiliation{Rudolf Peierls Centre for Theoretical Physics,  University of Oxford,Oxford OX1\ 3NP, United Kingdom}
\affiliation{SEPnet and Hubbard Theory Centre, Royal Holloway University of London, Egham TW20\ 0EX, United Kingdom}

\author{R. A. Borzi}

\affiliation{Instituto de Investigaciones Fisicoqu\'\i{}micas Te\'oricas y Aplicadas UNLP-CONICET, IFLP and Departamento de F\'\i{}sica, Universidad Nacional de La Plata, 1900 La Plata, Argentina}

\author{A. S. Gibbs}

\affiliation{SUPA, School of Physics and Astronomy, University of St Andrews, St Andrews KY16\ 9SS, United Kingdom}

\author{A. P. Mackenzie}

\affiliation{SUPA, School of Physics and Astronomy, University of St Andrews, St Andrews KY16\ 9SS, United Kingdom}

\author{R. Moessner}

\affiliation{Max Planck Institut f\"ur Physik komplexer Systeme,D-01187 Dresden, Germany}

\author{S. A. Grigera}
\affiliation{SUPA, School of Physics and Astronomy, University of St Andrews, St Andrews KY16\ 9SS, United Kingdom}
\affiliation{Instituto de F\'{\i}sica de L\'{\i}quidos y Sistemas Biol\'ogicos, UNLP-CONICET, La Plata 1900, Argentina}

\date{\today}

\begin{abstract}
We investigate the non-equilibrium behavior of the spin-ice material Dy$_2$Ti$_2$O$_7$ by studying its magnetization as a function of the rate at which an external field is swept. At temperatures below the enigmatic ``freezing'' temperature $T_{\rm equil}\approx600$~mK, we find that even the slowest sweeps fail to yield the equilibrium magnetization curve and instead give a smooth, initially much flatter curve. For higher sweep rates, the magnetization develops sharp steps  accompanied by similarly sharp peaks in the temperature of the sample. We ascribe the former behavior to the energy barriers encountered in the magnetization process, which proceeds via flipping of  spins on filaments traced out by the field-driven motion of the gapped, long-range interacting magnetic monopole excitations. In contrast, the peaks in temperature result from the released Zeeman energy  not being carried away efficiently into the bath, with the resulting heating triggering a chain reaction.

\end{abstract}
\maketitle
 
New phases go along with new excitations, which in turn manifest themselves as new phenomena in experiments. In the case of the magnetic pyrochlore compounds RE$_2$TM$_2$O$_7$, where RE = Dy or Ho and TM = Ti or Sn, the new phase, named \emph{spin ice}, is peculiar in that it is highly degenerate, with a non-vanishing ``zero-point entropy'' on account of the highly frustrated magnetic couplings \cite{gingrev} and an effective low-energy description provided by a classical emergent gauge field \cite{emgauge}.

The concomitant new excitations are -- unusually for a three-dimensional Ising magnet -- pointlike topological defects. These are charged not only under the emergent gauge field but are also sources of a conventional magnetic field, whence their name magnetic monopoles~\cite{Castelnovo2009}.  

The experiments we report and analyse here are related to the magnetization processes under an applied magnetic field. This is a fundamentally non-equilibrium phenomenon that has a long and important history in condensed matter physics. It is associated with concepts such as the Barkhausen noise~\cite{barkhausen}. It is also related to the physics of ``magnetic deflagration'' in the case of simple paramagnets~\cite{magdeflag}, with links to the physics of explosives, for which detailed analytic theories were proposed~\cite{Garanin2007}. 

In spin ice, we find that there are three fundamentally different types of behavior for the magnetization process. Besides the equilibrium magnetization curve for infinitesimally (at low temperature, $T$, unattainably) slow sweep rates, we find smooth behavior for slow sweeps, which gives way to  magnetization jumps accompanied by thermal runaway at faster sweeps.

This richness of behavior is a consequence of the peculiar nature of the spin ice state, which requires the motion of {\em thermally activated} monopoles to change the magnetization, leading to an exponentially slow response rate at low $T$. The slow-sweep response is special in that it reflects the nucleation of the equilibrium (magnetized) phase as in the case of a supercooled gas, with the crucial distinction that the nucleation happens through one-dimensional filaments (the paths of monopoles), so that the process can occur gradually even in the absence of quenched disorder. The relevant energy barriers arise from the competition of magnetic Coulomb and Zeeman energies involved in creating and separating the monopoles. In contrast, thermal runaway involves physics extrinsic to the spin system, namely a ``phonon bottleneck'', the inability of the heat bath to absorb efficiently the Zeeman energy released by the spin flips.

In the remainder of this paper, we describe and model these phenomena in detail. Specifically, we study the low temperature magnetization process in response to an external [111] field which takes place between two states within the ice-rule manifold: the zero field spin ice state (SI) and kagome ice (KI)~\cite{Matsushira2002}, defined below. 

The  pyrochlore lattice consists of a cubic array of corner-sharing
tetrahedra. In the spin ice systems, the spins -- sitting at the
corners of the tetrahedra -- are constrained to point into the center
of one of the two tetrahedra they belong to. The spin configurations in the low energy ice-rule manifold 
are those in which only two of the four spins of every tetrahedron point towards its center. Positive (negative) magnetic monopoles correspond to tetrahedra with three (one) spins pointing in.

To understand the transition between the zero field SI and KI states, it is convenient to view the pyrochlore lattice as an alternating stack of two-dimensional kagome and triangular lattice planes, perpendicular to the $[111]$ direction. It is possible to polarise all the spins in the triangular layers (more strongly coupled to the magnetic field on account of their local easy axes) while maintaining the ice rules and preserving a reduced but non-vanishing zero-point entropy~\cite{Matsushira2002}; this is the KI state. To further increase the magnetization along the $[111]$ direction requires breaking the ice-rules, with an energy penalty of $\Delta_s\approx 5.6$~Kelvin per spin flip. As a result, \Dy227 exhibits a low-temperature magnetization plateau at the value $M_{\rm KI} \approx 3.3$~$\mu_B$/Dy for fields up to about $1$~Tesla~\cite{Sakakibara2003}. On the other hand, at small fields the system's response is similar to that of an isotropic paramagnet and one expects a linear scaling of the magnetization with $H/T$~\cite{isakov}. 

Single crystals of \Dy227 were grown by the floating zone method in
St Andrews. The samples were cut into prisms of approx.
 $2 \times 0.7 \times 0.5$~mm$^3$ with the long axis along 
$[111]$ to minimize demagnetizing effects.  The magnetization was
measured using a purpose-built plastic Faraday force magnetometer.
The sample was mounted flat on a sapphire plate thermally anchored
to the mixing chamber of a dilution refrigerator. An additional calibrated
thermometer was attached directly to the sample. The magnetometer was calibrated against a SQUID.
%
\begin{figure}
\centerline{\includegraphics[width=0.85\columnwidth]{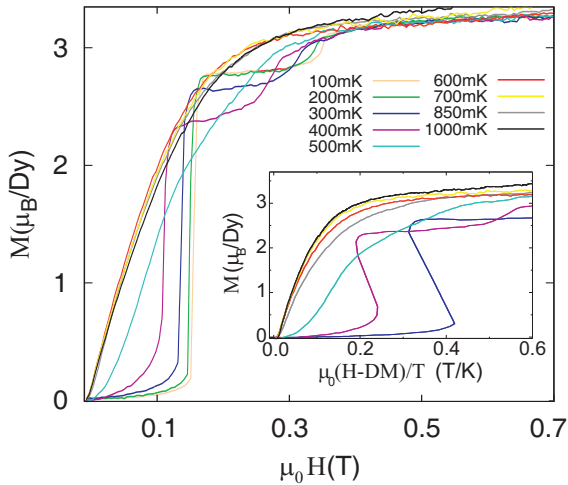}}
\caption{magnetization vs field along the $[111]$ axis, measured at low temperature after zero field cooling, at sweep rate $0.1$~T/min. While above $T_{\rm equil} \approx 600$~mK the spin system reaches equilibrium within the experimental measuring time (as shown by a reasonable $H_i/T$ scaling for low $H_i/T$, see inset), the lowest $T$ curves exhibit quite unusual features. Notably, plateaux are followed by sharp jumps in $M$. Eliminating the effect of demagnetizing fields, $DM$, gives a crucial piece of information (inset): the field triggers an event that persists even when the actual internal field $H_i$ decreases below the triggering value (the negative slope seen here is a consequence of the decrease in $H_i$ as $M$ increases). The details of the curves are remarkably reproducible.} 
\label{fig1}
\end{figure}

{\em Results --} Fig.~\ref{fig1} shows the measured magnetization along $[111]$ after zero field cooling, as a function of field (with a sweep rate $\upsilon =  0.1$~~T/m), for different temperatures. 
At relatively high temperatures (above $T_{\rm equil}$) the magnetization rises linearly from zero as expected and reaches a plateau of approximately $3.3$~$\mu_B$/Dy. As the temperature is lowered to $500$~mK, contrary to the expected $H/T$ scaling, the magnetization rises at a slower rate. Below $400$~mK the magnetization at first remains tiny until a finite field is reached, whereupon it  exhibits a steep jump which can hardly be resolved by our instrument, recording $1$ point every $\sim2$~s. 
Further plateaux at intermediate values of the magnetization and subsequent jumps are seen before reaching $M_{\rm KI}$. With decreasing temperatures, both the intermediate plateaux and the jumps become sharper and the value of the magnetization at the plateaux changes (from around $2.4$~$\mu_B$/Dy to $2.8$~$\mu_B$/Dy). The latter is a strong indication that the plateaux cannot be explained by the formation of an equilibrium intermediate magnetization pattern, in contrast to the case of magnetoelastic coupling~\cite{cabra}. 

Further information can be obtained by plotting the magnetization as a function of $H_i/T$, where $H_i = H-DM$, $D$ the calculated demagnetizing factor based on the geometry of the samples (inset of Fig.~\ref{fig1}). As expected, the $H_i/T$ scaling works only for the high temperature data, where it also agrees with MC simulations of the dipolar spin ice model (not shown). At lower temperatures the curves lie below the equilibrium curve. What is most striking is the fact that the jumps, plotted against the actual internal field, acquire a {\em negative} slope. 
This shows that they correspond to triggered events: once the process has started, it does not stop even though the internal field falls below the triggering value. This represents a magnetic Zeeman equivalent of 
deflagation of a combustible material set off by a spark.

\begin{figure}
\centerline{\includegraphics[width=0.85\columnwidth]{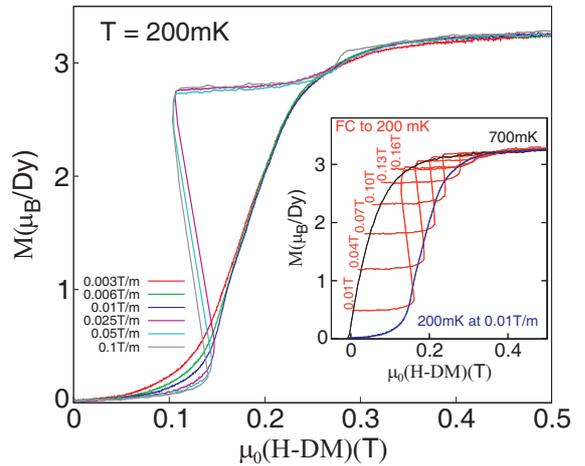}}
\caption{$M$ vs $H_i=H-DM$ at $200$~mK, measured for different field sweep rates $\upsilon$. Jumps only happen above $\upsilon = 0.025$~T/min. At slower sweep rates, the data seem to fall onto a single limiting curve. Jumps also occur in field cooling experiments; the inset shows $M$ vs $H_i$ at 0.1 T/min and 200mK for different starting field cooled states (as indicated in each curve), together with the slow curve (at 0.01 T/min) and the 700 mK curve.  The field-cooled jumps are contained within these two limiting curves, as explained in the text.} \label{fig2}
\end{figure}

Jumps in the magnetization have been observed indirectly in neutron studies of \Dy227\cite{fennell_jumps,f2} and were absent from the magnetization measurements reported in Ref.~\onlinecite{Sakakibara2003}. 
Fig.~\ref{fig2} reconciles this apparent inconsistency as a consequence of the different speeds at which the magnetic field was swept in the two experiments. The figure shows the magnetization curves at $200$~mK for sweep rates spanning a factor of about thirty, each step progressing by approximately a factor of two from the previous rate. They can clearly be separated into two groups: those showing one or more jumps, with sweep rates $\upsilon$ greater or equal to $0.025$~T/min, and those showing a continuous growth of the magnetization, from $0.01$~T/min to $0.003$~T/min. 
The changes within each group are very small, despite $\upsilon$ changing by a factor four: the heights of the jumps are unaltered and so is the slope of the continuous curve. 

The field at which $m$ jumps can be tuned by preparing the sample into different initial states of magnetization.  Shown
in the inset of Fig.~\ref{fig2} are $M$ vs $H_i$ curves at 0.1 T/min measured after cooling the sample to 200mK
under different applied magnetic fields; a slow curve (0.01T/min) at the same temperature and the 700 mK
curve are also included. We can see that the magnetization value for each field cooled curve hardly changes until the curve crosses the slow rate trace. Shortly after, the magnetization suddenly jumps up to its equilibrium value at $T \sim 700$~mK. 
The slow rate curve can thus be considered as an approximate
limit for the stability of the out-of-equilibrium magnetization.

\noindent {\em Low-temperature behavior} -- 
Focusing on field values below $0.1$~Tesla, we see that the magnetization at low temperature rises slowly, reaching higher values the smaller the sweep rate. Qualitatively, this behavior can be understood as a consequence of the kinetics of the magnetization process in spin ice, which proceeds via field-driven motion of the magnetic monopoles. 
As their density and speed of motion is not significantly affected by weak fields, the maximal rate at which the system can respond via this mechanism is independent of the value of the applied field: 
$(dM/dt)_{\rm max} \approx \rho_m \, 10 \mu_B/{\rm ms}$, where $\rho_m$ denotes the density of thermally activated unbound monopoles,  strongly suppressed at low temperature. Here we assume a typical single spin flip time scale of $1$~ms~\cite{Snyder2004,jaubert}. 
Even the slowest experimental sweeps require processes where $M$ changes much faster than $(dM/dt)_{\rm max}$ to maintain equilibrium for $T \ll T_{\rm equil}$. As a result, the system enters a strongly out of equilibrium regime where the magnetization remains very small despite the presence of an applied magnetic field.

When the sweep rate is sufficiently low, this regime is followed by a smooth, seemingly rate-independent increase in magnetization up to $M_{\rm KI}$ (Fig.~\ref{fig2}). Let us thus consider how monopoles, the agents of magnetization changes, are created out of the SI ground state in presence of a Zeeman field. 
On one hand, the bare cost of a spin flip, which creates a neighbouring monopole--antimonopole pair, $\Delta_s$, leads to a tiny density at low $T$, say 100 mK. On the other, a single monopole can change the magnetization by effecting $O(L)$ spin flips as it is swept to the sample surface, $L$ being the linear size of the system along the field direction.

We are thus dealing with a nucleation process where a spin flip out of the ground state is the part of an activated process akin to those giving rise to an interface tension between two phases in a first-order transition. The Coulomb attraction experienced by the monopoles as they move apart will then be reflected in the kinetics of the nucleation process.

We emphasize that -- unlike a supercooled gas, where a nucleated droplet grows to a finite volume fraction $O(L^3)$ -- a spin flip in SI only nucleates a filament of size $O(L)$, as mentioned above. In such nucleation processes, the limits of long times and large system sizes do not commute but it is easy to see that there will be a parameter range where the area density of filament creation is low enough for their Zeeman energy release to be absorbed by the rest of the system -- this is our slow-sweep regime~\cite{footnote}. As we will argue below, the fast-sweep regime corresponds to the situation where this is no longer the case, and thermal runaway is induced.

The smooth magnetization curve thus provides information about a sequence of energy barriers, which in principle depend on the field direction. In particular, its derivative $dM/dH$ at low temperatures is tantamount to a histogram of the distribution of such barriers.
A plausible origin for these barriers can be found in the Coulomb energy needed to separate a thermally excited monopole pair. Indeed, analytical considerations based on the effective monopole description of spin ice (as well as preliminary numerical simulations) yield energy scales that are in broad agreement with the experimental results for $dM/dH$ (not shown). However, a detailed microscopic modelling of this phenomenon is still needed to confirm the origin of the magnetization curve at low sweep rates. 
The $[111]$ field direction is particularly complicated as monopoles sweeping through the system can become stuck when their path encounters a spin in a triangular plane already pointing along the field. Such stuck monopoles give rise to local stray fields that  broaden the distribution of energy barriers around them. Moreover, spin flips in triangular and kagome planes have different Zeeman energies in a $[111]$ field. 

\noindent {\em Thermal runaway: the fast sweep regime} -- 
We now turn to the magnetization jumps, the defining distinction between the slow- and fast-sweep regimes. We believe that the increased rate at which Zeeman energy is dumped into the system as the field sweep rate increases overtaxes the ability of the lattice (the phonons) to equilibrate the system with the bath. As a result, the sample heats up locally, leading to the creation of more (and more easily unbound) monopoles. These in turn dump more energy as they move in the field direction and thermal runaway is `ignited' above a critical sweep rate. 

\begin{figure}
\centerline{\includegraphics[width=\columnwidth]{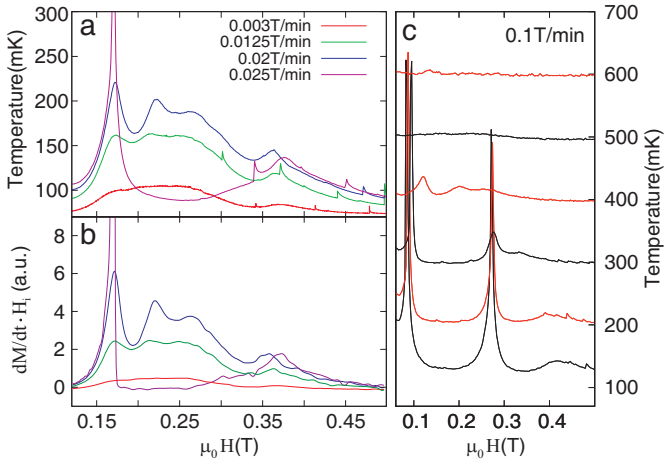}}
\caption{The temperature of the sample (panel a) follows closely the Zeeman energy release $\dot{W}_z$ due to the motion of monopoles in a magnetic field; the latter is proportional to $H_i(t) \, dM/dt$ (panel b). The temperature reached at each of the thermal runaway episodes is approximately independent of the sample initial temperature (panel c).}
\label{fig4}
\end{figure}

To verify this, we attached an additional thermometer directly to the sample to measure its temperature and magnetization simultaneously. 
In Fig.~\ref{fig4}(a) and~(b), we show this temperature for four different magnetic sweep rates as a function of field, along with $\dot{W}_z \propto H_i(t) \, dM/dt$, the rate of Zeeman energy release. 
The sample is strongly coupled to the bath (the dilution refrigerator mixing chamber) and therefore the local temperature $T_l$ should reflect $\dot{W}_z$. Indeed, $\dot{W}_z$ and $T_l$ trace each other closely. As seen in Fig.~\ref{fig4}, for slow sweep rates the energy is dissipated gradually during the whole magnetization process. Once the critical sweep rate is reached (about $0.025$~T/min in this case), the magnetization jumps abruptly, accompanied by a strong spike in the temperature. 

This picture is further supported by the observation that the maximal temperature reached by thermal runaway is approximately independent of the initial temperature over a wide range, pointing at an intrinsic feature of the thermal coupling in the spin-lattice system. This is displayed in Fig.~\ref{fig4}(c): at $T_{\rm equil} \approx$ 600 mK, thermal equilibration is efficient, resulting in a flat temperature trace. 
By contrast, for $T < 500$~mK temperature spikes appear which, crucially, never surpass $T_{\rm equil}$, where efficient thermal contact is clearly reestablished. This also explains why the magnetization always jumps up to roughly its equilibrium value at $T \approx T_{\rm equil}$ (inset of Fig. \ref{fig2}).

Precisely what it is that happens microscopically around $T_{\rm equil}$  is an important open question, which is all the more intriguing because another phenomenon is observed there: spin autocorrelations, as probed in numerical simulations {\em without phonons}~\cite{jaubert} and experimentally through AC-susceptibility, show an approximately exponential slowdown below $T_{\rm equil}$~\cite{Snyder2004}. 

In conclusion, we have found that the magnetization processes in spin ice exhibit a variety of novel out-of-equilibrium phenomena: nucleation of the equilibrium phase in one-dimensional filaments; thermal runaway following  `supercooling'; a distribution of energy barriers due to long-range interactions and a `speed-limit' on macroscopic equilibrium rearrangements resulting from a paucity of agents to effect such changes. These are exhibited robustly (i.e. without special preparation or a careful choice of parameters). Clearly, further studies of equilibrium and non-equilibrium dynamics and transport -- in particular, thermal transport -- are highly desirable\cite{footnote2}. 

We thank the Royal Society and EPSRC (UK), and CONICET and ANPCYT (Argentina) for financial support, and A. G. Green, T. S. Grigera, C. Hooley and S. Sondhi for useful discussions.


\begin{thebibliography}{}
\bibitem{footnote}Depending on nature and sequence of rate-limiting steps, the existence of a family of `slow-sweep regimes', as the sweep rate is varied over a broad range of values, would have been another possible scenario.
\bibitem{footnote2} Experimental measurements have shown that the nuclear spins in  Ho$_2$Ti$_2$O$_7$ have a far greater heat capacity than the electronic spins at low temperatures. It is likely that the nuclear spins act as a buffer in this compound, quickly absorbing the excess local heat released by electronic spin flip processes and transmitting it at a slower rate to the phonon bath. This would have the effect of quenching the runaway process up to much larger sweep rates/lower temperatures.
\end{thebibliography}
\end{document}